\documentclass[aps,prl,onecolumn,notitlepage,bibnotes]{revtex4-1}
\usepackage{amsmath,amssymb}
\usepackage{amsfonts}
\usepackage{relsize}
\usepackage{stmaryrd}
\usepackage{simpler-wick}
\usepackage{cancel}
\usepackage{mathtools}
\usepackage[T1]{fontenc}
\usepackage{geometry}
\newcommand{\no}[1]{:\mathrel{\mkern2mu #1 \mkern2mu}:}

\usepackage{hyperref}

\usepackage{bibentry}

\usepackage{natbib}
\usepackage{color}

\usepackage{graphicx}
\usepackage{dcolumn}
\usepackage{bm}
\usepackage{enumerate}
\usepackage{slashed}
\usepackage{comment}

\usepackage{soul}

\usepackage[english]{babel}
\usepackage{appendix}
\usepackage[utf8]{inputenc}

\usepackage{marginnote}
\usepackage[normalem]{ulem}

\usepackage{subfigure}

\newcommand*\xbar[1]{%
   \hbox{%
     \vbox{%
       \hrule height 0.5pt 
       \kern 1.0ex
       \hbox{%
         \kern-0.1em
         \ensuremath{#1}%
         \kern-0.1em
       }%
     }%
   }%
} 
\newcommand*\xxbar[1]{%
   \hbox{%
     \vbox{%
       \hrule height 0.5pt 
       \kern 0.8 ex
       \hbox{%
         \kern-0.1em
         \ensuremath{#1}%
         \kern-0.1em
       }%
     }%
   }%
} 

\newcommand{\uu}{{\rm{u}}}

\newcommand{\bz}{\bar z}

\newcommand{\ii}{{\rm{i}}}
\newcommand{\dd}{{\rm{d}}}
\newcommand{\vv}{{\rm {v}}}

\newcommand{\nn}{\nonumber}

\newcommand{\eq}[1]{(\ref{#1})}
\renewcommand{\>}{\rangle}
\newcommand{\la}{\label}
\newcommand{\ba}{\begin{align}}
\newcommand{\ee}{\end{equation}}
\newcommand{\be}{\begin{equation}}

\def\12{\frac{1}{2}}
\newcommand{\p}{\partial}
\newcommand{\en}{\end{align}}

\newcommand{\<}{\langle}

\usepackage{color}

\begin{document}

\setcounter{secnumdepth}{-1}

\title{ Quantization of Hydrodynamics: Rotating Superfluid, and  Gravitational Anomaly  }

\author{P. Wiegmann}
 \altaffiliation{also at IITP RAS, Moscow 127994, Russian Federation}
  \affiliation{Kadanoff Center for Theoretical Physics, University of Chicago,
5620 South Ellis Ave, Chicago, IL 60637, USA}

\date{\today}

\begin{abstract}
We present a consistent scheme of quantization of chiral flows (flows with extensive vorticity) of ideal hydrodynamics in two dimensions.   Chiral flows occur in rotating superfluid, rotating turbulence and also in electronic systems in  magnetic field in the regime of a fractional Hall effect. The quantization is based on a geometric relation of chiral flows to two-dimensional quantum gravity and is implemented by the gravitational anomaly. The effect of the gravitational anomaly violates the major property of classical hydrodynamics, the Helmholtz law: vortices are no longer frozen into the flow. Effects of quantization could be cast in the form of quantum stress. We show that the quantum stress generates Virasoro algebra, the centrally extended algebra of holomorphic diffeomorphisms.
\end{abstract}

\date{\today}

 \maketitle
\begin{flushright}{\it {Contribution for the JETP special issue in honor of  I.M. Khalatnikov centenary}}
\end{flushright}



\section{ Introduction}
The problem of quantization of hydrodynamics beyond linear approximation is commonly considered to be intractable.   Nevertheless, nature confronts us with beautiful quantum nonlinear  ideal fluids with experimentally accessible  precise quantization. Among them, two quantum fluids stand out: superfluid helium and electronic fluid in the fractional quantum Hall state. In both cases, the precise quantization of vortex circulation in superfluid helium and the precise quantization of electric transport in FQHE leave no doubts of the quantum nature of these fluids.

The fundamental aspects of quantization of fluid dynamics  most clearly appear in ideal flows. These are incompressible flows $$ {\bm \nabla}\!\cdot {\bm u}=0$$ of homogeneous   inviscid fluids.   

The problem of quantization  is further specialized in {\it chiral two-dimensional ideal flows}.  These are 2D flows  with  extensive  vorticity:  the mean vorticity 
\be 2\Omega=\frac 1 V\int \omega\, \dd V, \quad \omega(r)= {\bm\nabla}\!\times\!{\bm u}
\ee
remains finite, as $V\to\infty$. The chiral  flows are distinguished by holomorphic character of the quantum states.
 From the geometric perspective the quantization of ideal flows is equivalent
to a geometric quantization of  \(\bf SDiff\).

Two most perfect quantum fluids,  rotating superfluid helium \cite{Khalatnikov},
and FQHE fall to the class of ideal chiral flows (see \cite{wiegmann2013hydrodynamics,wiegmann2013JETP}
for the correspondence between FQHE\ and superfluid hydrodynamics).  The superfluid helium is a compressible fluid, however, rotating superfluid  could be considered as incompressible. In a good approximation its flows are two-dimensional. These are flows of vortex cores in the plane normal to the axis of rotation.  States of electronic fluid in FQH regime, where electrons fractionally occupy the lowest Landau level are holomorphic, hence their flows are incompressible. 

In this paper we show how to quantize chiral flows and describe some, not immediately obvious, consequences of  quantization. The guidance for the quantization comes from    the intersection between quantum chiral hydrodynamics   and quantum two-dimensional  gravity.
The relation between these two subjects is  briefly described here.

Ideal flows are Hamiltonian. The    Hamiltonian is the kinetic energy of the flow
\begin{align}\nn
&H=\frac{\rho_{\scriptscriptstyle\rm A}}2\int \bm u^2\dd V
\end{align}
where \(\rho_{\scriptscriptstyle\rm A}\) is a constant density of the fluid.
The Poisson structure is  the Lie-Poisson algebra of area preserving diffeomorphisms    \(\bf SDiff\)\begin{align}\la{6}
&\{\omega({ r)},\ {\omega( r'})\}=\rho_{\scriptscriptstyle\rm A}^{-1}\left(\nabla_{
r}\times\nabla_{
r'}\right)\omega({ r)}\delta ({ r- r')} .
\end{align}
  Hence flows of ideal fluid  are   the actions of area-preserving diffeomorphism \(\bf SDiff\), and   should be studied from a geometric standpoint, see e.g., \cite{arnold1999topological}.

Formally the quantization amounts  supplanting  the Poisson brackets  by the commutator \(\{\ ,\ \}\to\frac
1{\ii\hbar}[\ ,\
]\)  (first considered in Landau's 1941 paper   \cite {Landau1941}), and  identifying   the Hilbert space with representation space of 
\(\bf SDiff\).      The latter, however, is not well understood.  As always in quantization of continuous media, difficulties appear  in regularization of  short-distance  divergencies.   A  regularization must be consistent with fundamental symmetries of the  theory, especially with local symmetries.   The  local   ``symmetry'' of  fluids is the  relabeling symmetry,  or equivalently the invariance  with respect to  reparametrization  of  flows.  In this paper we describe the diffeomorphisms invariant  regularization and show that it alone dictates the universal quantum corrections to the Euler equation.

Relabeling  are  diffeomorphisms in the manifold of Lagrangian coordinates.  Invariance with respect to diffeomorphisms is also a guiding principle of quantum gravity. In the paper we describe the correspondence between chiral flows and 2D gravity,  and show how  this correspondence yields to a unique   regularization.  The result of the regularization   is quantum corrections to the Euler equation expressed in terms of the gravitational anomaly.
These corrections are small in superfluid helium,  and may or may not be negligible in cooled atomic gases. But they are certainly large and observable  in FQH electronic fluids. But regardless of their size quantum corrections are fundamentally important for consistency of the theory.

For the purpose of this paper, we choose the quantization within the  Eulerian specification.  Other methods of quantization, such as, via path integral and stochastic quantization yield the same result and will be published separately.

\section{ quantization in the Lagrangian specification}In this paper we consider only the bulk flow,
ignoring complications caused by boundaries and multiply connected geometry.  The bulk incompressible flows solely described by vorticity.  The quantization scheme we propose consists of two  independent steps.

\subsection{Semiclassical quantization}

The first step is the quantization of vortex circulation. In quantum ideal fluids vorticity is not  a smooth function. It consists of a discrete array of  point-vortices \begin{align}
\omega( r)\equiv {\bm\nabla}\times{\bm u}=\sum_{i=1}^{N{\scriptscriptstyle\rm v}}\Gamma_i\delta( r- r_i),\la{1}
\end{align} 
whose    circulation   \(\Gamma_i\)  is quantized in units of \(h/m_{\scriptscriptstyle\rm A}\) (\(m_{\scriptscriptstyle\rm
A}\) is the mass of fluid atoms).

 Hence, quantum flows confined in a finite volume are always finite dimensional. If we stop here we face a problem of classical
vortex dynamics.  We  formulate the dynamics in the form of Kirchhoff. 
\subsection{Kirchhoff equations}

We recall   the    Helmholtz  form
of the  hydrodynamics of ideal  2D  flows.  The Helmholtz law is the curl of Euler equations.  It stays that vorticity is frozen into the flow:  the  {\it material derivative of vorticity
 vanishes} \begin{align}
 \quad D_t\,\omega=0,\quad D_t=\p_t+\bm u\cdot\bm\nabla. \la{2}
\end{align}
 On other hand the flow itself is determined by positions of vortices. The complex   velocity \(\uu=u_x-\ii u_y\) of  an array of \(N_\vv\) vortices \eq{1} is a meromorphic  function. On a Riemann sphere in
complex coordinates \(z,\bz\) it reads \begin{align}
\uu(z)=\frac\ii{2\pi}\sum_{i=1}^{N_\vv}\frac{\Gamma_i}{z-z_i}.\la{3}
\end{align}     
Then the Helmholtz law and Kelvin circulation theorems say that the flow initially chosen as \eq{3}  retains its form with positions of vortices \(z_i(t)\) moving accordingly the Kirchhoff equation (see, e.g. \cite{saffman1992vortex})
\begin{equation}
\dot{\bar{z}}_{i}=\frac \ii{2\pi}\sum_{j\neq i}\frac{\Gamma_{j}}{z_{i}-z_{j}}.\la{4}
\end{equation} 
Kirchhoff equations, being exact, for flows with a finite number of vortices
 ``approximate''   arbitrary ideal  flows  by a finite dimensional   dynamical
system. However, in the quantum case they are not an approximation. Rather,
they are starting point of quantization.

Kirchhoff equations are Hamiltonian. The   Hamiltonian and the Poisson brackets are
\begin{align}\la{5}\begin{split}
&H=-\frac{\rho_{\scriptscriptstyle\rm A}}{2\pi}\sum_{i> j}\Gamma_i\Gamma_j\log|z_i-z_j|, \\
&\{ z_i,\ \bz_j\}=2\ii(\rho_{\scriptscriptstyle \rm A}\Gamma_i)^{-1}\delta_{ij}.
\end{split}
\end{align} 
      Hence  the    dynamics of a finite number of vortices could be quantized canonically.  
This is the second step of quantization.
In this approach trajectories  of vortices
\(z_i\), rather than fluid atoms, are pathlines of (vortex) fluid.  

In chiral flows when the number of vortices tends to infinity vortices themselves form a fluid. Then 
enumeration  of vortices, the label \(i\) is the    
  Lagrangian
coordinate of the vortex fluid.

\subsection{Canonical quantization}
Within canonical quantization we  identify the Hilbert space with the space   of meromorphic  functions of  coordinates of clockwise and anti-clockwise vortices,   map the coordinates to operators acting in the space of vortices and  supplant the Poisson brackets by the commutator \(\{ z_i,\ \bz_j\}\to\frac 1{\ii\hbar}[z_i,\
\bz_j]\). Since the vortex circulation is  quantized  in units of \(\Gamma =h /m_{\scriptscriptstyle\rm A}\),  and \(\rho_{\scriptscriptstyle\rm A}=m_{\scriptscriptstyle\rm A} n_{\scriptscriptstyle\rm A}\),  where
  \(n_{\scriptscriptstyle\rm A}\) is the density of fluid atoms,
 the commutation relations read\begin{align}\nn
[ z_i,\ \bz_j^{}]=-{\rm sign}(\Gamma_i)(\pi  n_{\scriptscriptstyle\rm A})^{-1}\delta_{ij}.
\end{align}
This scheme is a guidance for the quantization of hydrodynamics. It differs from earlier quantization attempts where flow was described by pathlines of fluid atoms.

 If the number of vortices is finite the quantization presents no fundamental difficulties. The difficulties arise when we attempt the large \(N_{\rm v}\) limit when an array of vortices approximate realistic flows of interest. Among them are chiral flows we consider below.

\subsection{Chiral flows}

The 2D chiral flows are flows with extensive net vorticity (vorticity is proportional to the volume of the fluid). Such flows consist of  a dense liquid array of vortices.  We assume that circulations of all vortices are the same \(\Gamma_i=\Gamma>0 \) (anticlockwise). Then vorticity is proportional to the vortex density \begin{align}
   n( r)=\Gamma^{-1}\omega(r)=\sum_{i=1}^{N{\scriptscriptstyle\rm v}}\delta( r- r_i).\la{71}  \end{align} 
 We denote the mean density of vortices  by 
\(n_\vv=N_\vv/V,  \)   the density of  fluid atoms
\(n_{\scriptscriptstyle\rm A}=\rho_{\scriptscriptstyle\rm A}/m_{\scriptscriptstyle\rm A}\) and the fraction \(\nu=N_{\scriptstyle \rm v}/N_{\scriptscriptstyle\rm A}\) of vortices per fluid atom. We also use \(m_\vv=\nu^{-1}m_{\scriptscriptstyle\rm A}\), the ``roton'' mass, the mass of fluid per vortex.

The well-known examples of quantum chiral 
flows are  superfluid helium rotating with the angular frequency \(\Omega=
\frac\Gamma 2  n_\vv\),
and also   FQHE, where \(N_\vv\)  electrons confined in a 2D layer are placed in a uniform magnetic field   \(B=(h/ e)n_{\scriptscriptstyle\rm A}\), whose total number of flux quanta \(N_{\scriptscriptstyle\rm A}=(e/h)BV\) equals the number 
of fluid atoms.  In this case, electrons occupy a fraction \(\nu\) of the lowest Landau level.  

The chiral flow sets the scale of length, the mean distance between vortices \(\ell_\vv= (2\pi n_\vv)^{-1/2}\), and \(\hbar\Omega\) sets the energy scale.
We collect all these units below
\begin{align}\nn\begin{split}
&n_{\scriptscriptstyle\rm A}=\frac{\rho_{\scriptscriptstyle\rm A}}{m_{\scriptscriptstyle\rm A}} =\frac e h B, \quad \Gamma=\frac h{m_{\scriptscriptstyle\rm A}},     \\&  n_{\vv}=\frac{2\Omega}\Gamma=\frac 1{2\pi\ell_\vv^2},\quad  \nu=\frac{n_\vv}{n_{\scriptscriptstyle\rm A}},\\& m_\vv=\nu^{-1}m_{\scriptscriptstyle\rm A},\quad \hbar\Omega=\frac{\hbar^2}{2m_{\scriptscriptstyle\rm A}\ell_\vv^2}. \end{split}
\end{align}In these
units the Kirchhoff Hamiltonian and  the commutation relations  read
 \begin{align}\begin{split}
&H=-\frac{\hbar\Omega}{\nu}\sum_{i\neq j}\log|z_i-z_j|,\\ & [z_i,\bz_j]=(\pi n_{\scriptscriptstyle\rm A})^{-1}\delta_{ij}.\la{7}
\end{split}
\end{align} In its turn the  Kirchhoff equations \eq{4} become  Heizenberg equations
for vortices
\begin{equation}\nn
\ii \hbar\dot{\bar{z}}_{i}=(\pi n_{\scriptscriptstyle\rm A})^{-1}\partial_{z_i} H=-\frac \hbar{2\pi} \sum_{j\neq i}\frac{ \Gamma_{}}{z_{i}-z_{j}}.
\end{equation}

The fraction \(\nu\) is de facto a semiclassical parameter. It is small in helium, but is of the order one, say \(\nu=1/3\) in FQHE. 

The  Hilbert space of the chiral flow, where  vortices are of the  same sense,   is the Bargmann space \cite{girvin1986,bargmann1961}. It is a space of holomorphic
polynomials of \(z_i\) and  \(\p_{z_i}\)   with the inner product \be\<g|f\>=\int
\ g( \bar z)f(z)\dd \mu ,\quad \dd\mu=e^{-\pi {n_{\scriptscriptstyle\rm A}}|z|^2}\dd z \dd\bz. \nn\ee  Then the anti-holomorphic coordinates 
operators \(\bar z_i\)   are    Hermitan conjugations of   holomorphic coordinates
\(z_i\)  realized as     \be\nn\bz_i=z_i^\dag,\quad z_i^\dag=-(\pi n_{
\scriptscriptstyle\rm A})^{-1}\overset{{}_{\shortleftarrow}}{\p}_{z_i},\quad  z_i^\ast\equiv(z_i^\dag)^{\scriptscriptstyle\rm T}=(\pi n_{\rm
\scriptscriptstyle\rm A})^{-1}\vec\p_{z_i}.\ee
 All holomorphic states are  eliminated by \(z_i^\dag\) 
 \begin{align}\nn
\<\text{\ holomorphic  states}\ |z_i^\dag=0.
\end{align}
We can easily determine the ground state wave function of the chiral flow. It is  the stationary flow corresponding to a solid rotation of the vortex system \(z_i(t)|0\>=e^{-\ii\Omega t}z_i(0)|0\>\). In this case the velocity \(\dot{\bar{z}}_{i}|0\>=-\ii\Omega\bz_i|0\>=\nu(\Gamma/2\pi\ii)\p_{z_i}|0\>\). Hence, the ground state is nulled by the operator\begin{align}\nn
\begin{split}&\left(\hbar\Omega\p_{z_i} +\partial_{z_i} H\right)|0\>=0,\;\text{or}\;\\& \left(\nu\p_{z_i} -\sum_{j\neq i}\frac{1_{}}{z_{i}-z_{j}}\right)|0\>=0.
\end{split}
\end{align}
This equation has single valued solutions only if \(\nu^{-1}\in\mathbb Z\) is integer: each vortex is surrounded by the integer number of atoms. This is well known quantization of the inverse fraction in FQHE. Then the stationary   state wave function  is the holomorphic polynomial \begin{align}
|0\>_B=\prod_{i>j}(z_i-z_j)^{1/\nu}.\la{14}
\end{align}
This is of course  the  Laughllin wave function for FQHE  \cite{Laughlin1983} in the  Bargmann space.

If we choose the standard $L^2$  scalar product, the factor \(e^{-\frac 1{2}\pi n_{\scriptscriptstyle\rm A}\sum_i|z_i|^2}\) must be added to the holomorphic polynomial \eq{14}.
In this case the Laughlin wave function  of a stationary chiral  flow appears in a more familiar form \begin{align}
|0\>_{L^2}=e^{-\frac 1{2}\pi n_{\scriptscriptstyle\rm A}\sum_i |z_i|^2}
\prod_{i>j}(z_i-z_j)^{1/\nu}.\la{161}
\end{align}
The probability distribution of vortices \(\mathcal{\dd P }=||0\>|^2{\dd V}\) in the ground state \eq{161} could be  expressed through the Kirchhoff free energy \(H-\Omega L\)   as  \begin{align}
||0\>|^2=e^{-\frac 1{T_*} (H-\Omega L)},\quad T_*=\hbar\Omega,
\la{16}
\end{align}
where 
\be
L=m_\vv {\sum_i\bm r_i}\times{\bm v_i}\la{171}
\ee
 is
the angular momentum  of the vortex matter, also called angular impulse \cite{saffman1992vortex},    and \(\bm v_i=\dot{\bm r_i}-\Omega\times \bm r_i\),
the velocity of a vortex in rotating frame.

This formula looks as a canonical Gibbs  distribution with the temperature \(T_*=\hbar\Omega\).     This ensemble appears in various independent contexts and  had been extensively studied (see, e.g., \cite{Zabrodin2006}). Among them are  one-component plasma, Dyson diffusion, and also Onsager ensemble of vortices.     It is also the equilibrium distribution of   a non-determinantal stochastic point process,  called \(\beta\)-ensemble, where \(\beta\)  labels \(\nu^{-1}\).
 At large \(N_\vv\) it describes a distribution with an equilibrium density of vortices equal \(n_\vv=2\Omega/\Gamma\) \cite{FEYNMAN1955}, supported by a disc of the area \(V=N_\vv/n_\vv\).
 \section{LIe Algebra of Area preserving diffeomorphisms}
It is a straightforward check  that the brackets \eq{5} for vortex coordinates with the relations
(\ref{1}) yield the  Poisson structure for the ideal fluid
\eq{6},
\textbf{SDiff}. Hence, the quantum version of  \textbf{SDiff} could be also realized by operators \(z_i,\bz_i\) with the algebra \eq{7} acting in the Bargmann  space. 

For the  remaining  part of the paper we assume the torus geometry. Consider  Fourier mode expansion of the vortex occupation  number \eq{71} \(n_k=\int e^{-\ii {\bm k\cdot \bm r} }n({ r})\,\dd V=\sum_i e^{-\frac \ii 2( {\rm k}\bz_i+\bar {\rm k} z_i)}\). In the Bargmann space the occupation number is  realized by   the normally ordered  operator with respect to  holomorphic  states\begin{align}
 {\no{ n_k}}=\sum_{i\leq N_\vv}e^{-\frac{\ii}
2{\rm
k}z_i^\dag}e^{-\frac{\ii} 2{\rm
\bar k}z_i},\nn
\end{align}  where \({\rm k}=k_x+\ii k_y\) is a complex wave vector.   It gives the quantum meaning to vorticity \(\no{\omega(r)}=\Gamma V^{-1}\sum_k e^{\ii\bm k\cdot\bm r}\no{ n_{k}} \).

  The occupation number  operator is chiral \(n_{k}^\dag=  n_{-k}\), and eliminates the ground state, unless \(k=0\)
\begin{align}
   \<0|\no{  n_k}=\<0|N_\vv \delta_{k0}.\la{202}
\end{align}With the help of \eq{7} we  obtain  their algebra \begin{align}
\begin{split}\la{15}
&[\no{ n_k},\ \no{ n_{k'}}]=\ii e^{ \left( \frac{\bm k\cdot \bm k'}{4\pi n_{\scriptscriptstyle\rm A}}\right)} e_{kk'} \no{n_{ k+ k'}}\\ &\end{split}
\end{align}  
with the structure constants   
\begin{align}     e_{kk'}=2\sin\left( \frac{\bm k\times{\bm
k}'}{4\pi n_{\scriptscriptstyle\rm A}}\right).\end{align}
On a torus, where \((L_{1,2}/2\pi) k_{1,2}\in \mathbb{Z}\) is integer, the algebra is finite dimensional.

  We comment that for certain physical problems  not  ordered operators \(
n_{k}=\sum_i e^{-\frac \ii
2( {\rm k}\bz_i+\bar {\rm k} z_i)}=e^{(8\pi n_{\scriptscriptstyle\rm A})^{-1}k^2}\no {
 n_{k}}
\) could
be relevant.  Their   algebra is \([ n_{k},\
 n_{{k}'}]=\ii e_{kk'} n_{ 
k+ k'}\).  It is also known
as  {\it sine-algebra}
\cite{fairlie1989infinite}. The sine-algebra had been introduced as a finite dimensional  ``approximation'' of \textbf{SDiff}.  As we had seen this algebra naturally appears as a result of quantization. 

   In the  classical limit,   and  in the long wave limit when we   neglect  the discreetness of atoms the structure constants   \begin{align}
e_{kk'}\approx (2\pi n_{\scriptscriptstyle\rm A} )^{-1}\left(\bm k\times{\bm k}'\right).\la{172}
\end{align}      bring us back to the  Lie-Poisson  structure  of  classical
hydrodynamics
 \eq{6}.   
\section{Quantization in the Eulerian  specification}
Despite of straightforward quantization of   a finite number of vortices described above,  the   quantization of  flows approximated by a  large number of vortices met essential difficulties.
The source of difficulties is the same as in any quantum theory of continuous media  - a passage from the finite dimensional to the infinite dimensional system.  Formally the problem arises as follows.  The advection term $({\bm u}\cdot\bm\nabla)\omega$ in the Helmholtz equation \eq{2} possesses two operators sitting at the same point. This requires a regularization.
 A standard recipe of a  regularization commonly adopted in the field theory is point splitting ${\bm u}({ r}+\frac{\ell_\vv}
2)\cdot \bm\nabla \omega({ r}-\tfrac{\ell_\vv} 2)$, when points are split by a shortest distance between atoms. However, in  hydrodynamics, this recipe is inconsistent with the  relabeling symmetry of the fluid.  The difficulty is that contrary to a field theory,  the point-splitting distance, the short distance cut-off in hydrodynamics is variable. It itself depends
on the flow.   Similar effect is known
in the quantization of gravity \cite{Polyakov1987}.     In the nutshell, the idea is that a  variable   short-distance
cutoff is the distance between  vortices  \(\ell[n]\sim  n^{-1/2}\).
In the superfluid theory this regularization 
had been explored  a long ago. It has been used  for regularization of vortex energy in rotating superfluid and explained in details in Khalatnikov's book  \cite{Khalatnikov}, see also original papers \cite{bekarevich,kemoklidze,hall1960,hall-vinen}. Corrections to the energy of the flow obtained in these papers take into account the discreteness of vortices, otherwise have a classical nature and locally depend on vorticity.  These corrections are Casimir functions. They contribute to the stress of the vortex fluid (nowday they are called odd or anomalous stress \cite{AW,bogatskiy2019vortex}), but because they are local that stress is divergent-free.    It does not affect the  Euler equations for the bulk flow and only essential on  curved surfaces \cite{bogatskiy2019vortex}, at the boundary   \cite{kemoklidze}, and at the    interface between the vortex array and a potential flow \cite{bogatskiyedge}.   
 
Here we push this idea forward. We determine the quantum correction. The quantum correction depends on a gradient of vorticity, hence enters the equation  for the bulk flow.  We express it in terms of a quantum stress. 
\subsection{Quantum corrections to the Helpmholtz equation and quantum stress}
The quantization of the field equations, such as Helmholtz equation, essentially
means normal ordering of operators entered the   equation with
respect to a flow of interest.  The
regularization problem,  discussed above,  comes into effect   when we  attempt
to order  the  advection
term\begin{align}
({\bm u}\cdot{\bm \nabla})\omega=\epsilon^{ik}\p_k\nabla^j \left(u_i u_j\right).\nn
\end{align}  Hence, we have to understand
the quantum meaning of
 the traceless part of  the momentum flux tensor \(\rho_{\scriptscriptstyle\rm A}u_iu_j\). It requires to order  the product of two velocity  operators.  
  We  denote the Wick contraction  \(\wick[offset=0.8em]{\c A\c B}=\no {AB}\!-\!\no{
A}\no{ B}\) and compute the traceless part of   \(\wick[offset=0.6em]{\c u_i\c u_j}\).  This is the  quantum
correction received by the momentum flux tensor. It  should be interpreted
as a stress. We follow the notations and terminology of Landau and Lifshitz hydrodynamics text book   \cite{Landau6} (notice the
minus sign in the definition)\begin{align}
\begin{split}&
\rho_{\scriptscriptstyle\rm A}\no{u_iu_j}=\rho_{\scriptscriptstyle\rm A}\no{u_i}\no{u_j}\!-T_{ij},\quad T_{ij}=-\rho_{\scriptscriptstyle\rm A}\wick[offset=0.65em]{\c
u_i\c u_j}. \la{362}
\end{split}
\end{align} 
 Notice that only the traceless part of the stress enters the Helmholtz equation.  Latter (see (\ref{31},\ref{36})) we define and determine  the trace of the stress as forces  exerted upon  vortices. It differs from the  force exerted on the fluid, and, in particular does not involve pressure.

   Since quantization is the sole origin of the stress we refer it as a {\it
quantum stress}. The quantum stress corrects  the Euler equation and the  Helmholtz law 
 \begin{align}
 \begin{split}
 &\rho_{\scriptscriptstyle\rm A}D_t u_i+\nabla_i p=\nabla^j {
T}_{ij},\\
 &\rho_{\scriptscriptstyle\rm A}D_t\omega=\epsilon^{ik}\p_k\nabla^j {
T}_{ij}. 
\end{split}\la{27}
\end{align}
In this equation, all entries are assumed to be normally ordered, and so the equation could be treated as classical (e.g., \(D_t=\p_t+\langle\bm
u\rangle\cdot\bm\nabla\). The implication of quantum stress is that  the major property of classical
hydrodynamics,  the Helmholtz law, held  for quantum operators  does not
hold for their matrix elements. We comment that the stress has other terms originated by the effect of excluded volume, known as odd-viscosity \cite{Khalatnikov,AW}. However this term is divergence free and does not enter into the bulk equation, but affects the boundary \cite{kemoklidze,bogatskiyedge}. We do not consider it here.

The quantum stress  corrects the Helmholtz law.
Vorticity is no longer frozen into the flow. Nevertheless the Kelvin theorem (conservation of vorticity of a fluid parcel is intact). The Kelvin theorem  holds when  the divergence of the stress has no circulation along a liquid contour. It, indeed, vanishes. The formulas    (\ref{31},\ref{36}) obtained below yield  
\(\oint \nabla^j {T}_{ij}\dd x^i=\12\oint T^i_{\ i}\dd n\).   The latter  obviously vanishes.

We  will show that   the quantum  stress   reads
 \begin{align}
 \begin{split}
T_{ij}&=\frac {\hbar\Omega}{96\pi}\Big[\left(\nabla_i\nabla_j-\delta_{ij}\Delta\right)\log n
-\left(\nabla_i\log n\,\nabla_j\log n-\12\delta_{ij}\left(\nabla\log n\right)^2\right)\Big].\nn
\end{split}
\end{align}
This formula holds when vorticity is a smooth unction of coordinates and is everywhere positive. For references we write the stress   in complex coordinates\begin{align}
&T_{ij}\dd x^i\dd x^j=\frac
14[T_{zz}(\dd z)^2+2T_{z\bz}\dd z\dd\bz+T_{\bz\bz}(\dd
\bz)^2].\nn\\
&T_{zz}=\frac{\hbar\Omega}{12\pi} \left( \partial_z^{2}\log n-
\frac{1}{2}(\partial_z\log n)^{2}\right),\la{271}\\
&T_{z\bz}=-
\frac {\hbar\Omega}{12\pi} \, \p_z\p_{\bz}\log n,\quad T_z^{\bz}=nT_{zz}.\nn
\end{align} The Helmholtz equation then reads\begin{align}
\rho_{\scriptscriptstyle\rm A}D_t\omega=\12\nabla T_z^{\bz}\times\nabla\omega
\end{align}The quantum stress is 
 small. It   consists of  higher derivatives, 
but it is the only source which   deviates  vortices   away from   the flow.

\bigskip

Below we obtain these formulas.
\section{Geometric interpretation of the chiral  flows as  quantum gravity}

We start from a general discussion of   parallel between   2D chiral flow 
and 2D gravity.
The surface which hosts the fluid is a complex manifold equipped with a closed vorticity 2-form,  \(\omega_{ij}\dd x^i\wedge\dd x^i\), \(\omega_{ij}=\p_i u_j-\p_ju_i\). Because vorticity of the chiral flow does not change sign, in our convection, it is positive \(\omega=\12\epsilon^{ij}\omega_{ij}=\Gamma n>0\),    the chiral flow gives the host surface a K\"ahler structure with the K\"ahler form \(n \dd z\wedge\dd\bar z\), and the Riemannian metric $$\dd s^2=n|\dd z|^2.$$  The volume element of the metric   \(\dd N_\vv=n\dd V\) is  the number of vortices  in  the fluid    volume  \(\dd V\).  Hence, we can associate the chiral flow with an auxiliary
surface with the metric \(\dd s^2\) and  interpret fluid  flows  as  flows of the metric. 
    Adopting the language of quantum gravity we may identify the manifold
of  Lagrangian coordinates with a target space, and treat the host surface
 as a world-sheet. We comment that the analog of the metric \(\dd s^2\) 
had been used in studies of the vortex lattice
in \cite{andreev1984,volovik1980}.
Here we utilize it for liquids. 

     The map of  the coordinate frame
to the tangent space of the auxiliary surface   \((x^1,x^2)\to (\lambda^1,\lambda^2)\)   is realized through  Clebsch variables (see, e.g., \cite{zakharov1989algebra}). We recall that Clebsch variables  parameterize vorticity as \be n= \nabla\lambda^    1\times \nabla\lambda^2,\la{Cl}\ee and, that the intersection of level lines of \(\lambda^1\) and \(\lambda^2\) are position of vortices.  Hence the Clebsch variables  could be interpreted as  vortex labels. Hence, vorticity is  the Jacobian
of the map \(n={\rm  det}\|\nabla_i\lambda^a\|\), and   \(e^a_i\dd x^i=\dd \lambda^a\) are the vielbeins and  the metric \(\dd s^2=\dd\lambda^a\dd \lambda^a\).
The diffeomorphisms in the space of Clebsch variables which does not change the vorticity  is  relabeling of  vortices. A choice of Clebsch variables \(\dd\lambda^1+\ii\dd\lambda^2=\dd z/(\sqrt{2\pi}\ell[n])\), where \(\ell[n]=(2\pi n)^{-1/2}\) is of the order of a distance between vortices brings the metric to the diagonal form \(\dd s^2=|\dd z|^2/\ell^2[n]=n|\dd z|^2\).
Then the interval \(\dd s=|\dd z|/(\sqrt{2\pi}\ell[n])\)  is the ``distance" between vortex labels of vortices separated by the physical distance \(|\dd z|\).
If the fluid resides on a flat surface, then, generally, the  auxiliary surface  is curved. Its   scalar curvature is
\begin{align}
\mathcal{R}=-(4/n)\p_z\p_{\bz}\log n.\la{101} 
\end{align}
  A re-parameterization of the axillary surface is equivalent to relabeling vortices with no effect on the flow. This is   relabeling symmetry, or  a diffeomorphism invariance of  hydrodynamics. 
The relabeling symmetry usually refers
to fluid atoms. In our approach, it is a relabeling symmetry of vortices.
We want to keep this major symmetry intact in quantization. This amounts to implement a uniform short-distance cut-off on the target space (the space of labels). In the host plane (the world-sheet) the cut-off is a function of the flow.
This principle uniquely determines the regularization.

We illustrate this idea in terms of the path integral approach to quantization. In this approach, one  integrates over pathlines of fluid parcels. Instead, we choose to integrate over pathlines of vortices. If the number of vortices is finite, the measure of the path integral  is trivially invariant with respect to   relabeling. However, we have to integrate over flows which consist of a large number of vortices and in the coarse-grained limit when vorticity  could be approximated by a smooth function. In this case the relabeling does  not  hold  trivially. It determines the measure of the vorticity. In order to   determine it we invoke the relation of the chiral flow to 2D quantum gravity.   Since vorticity is a metric  we effectively   integrate over     metrics \(\dd s^2 =n|\dd z|^2\).  The measure in the space of metrics   has been established in quantum gravity \cite{Polyakov1987}. It consists of the Fadeev-Popov determinant restoring the re-parametrization invariance of the surface. In quantum gravity it is a source of the Liouville action. In quantum hydrodynamics, it  is a source of quantum stress. 

Below  we present a different  but  economic
approach to the quantization based on diffeomorphism  invariance.
In practice, it requires that all observable quantities are expressed through invariant geometric objects of the metric $\dd s^2$, such as geodesic distance, curvature, etc.

\section{Gravitational anomaly} 
We now describe the major 
effect of quantization, the gravitational anomaly.  In complex coordinates     \(
\uu_z=u_x-\ii u_y
\  \p_z=\12(\p_x-\ii\p_y)  \)
 the  advection term and the quantum stress read 
\begin{align}
\no{({\bm u}\cdot{\bm \nabla})\omega}=\rho_{\scriptscriptstyle\rm A}^{-1} \ii [\p_z^2 T_{\bz\bz}-\p_{z}^2T_{\bz\bz}],\quad T_{zz}=-\rho_{\scriptscriptstyle\rm A}\wick[offset=0.65em]{\c
\uu_z\c \uu_z}.\la{53}
\end{align}
 We express the stress through the occupation number with the help of the formula \(\uu_z=2\ii \Gamma\p_zV^{-1}\sum_k
e^{\ii\bm k\cdot\bm r}k^{-2}n_k\)  
\begin{align}\nn\begin{split}
\wick[offset=0.65em]{\c
\uu_z\c \uu_z}=-& \Gamma^2\underset{r\to r'}\lim(4\p_{z'}\p_z)
\frac 1{V^2}\sum_{k,k'}e^{\ii(\bm k\cdot \bm r+\bm k'\cdot \bm r')}(k'k)^{-2}\wick[offset=0.65em]{\c
n_k\c n_{k'}}.\end{split}
\end{align} Hence  we need to compute  the  contraction of the normally ordered      two-modes operator. Their normal ordering reads
\begin{align}\no{n_{ k}n_{ k'}}\ =\sum_{i,j}
e^{-\frac\ii 2{\rm
k}z_i^\dag}e^{-\frac\ii 2{\rm
k'}z_j^\dag}e^{-\frac \ii 2\bar{\rm
k}z_i}e^{-\frac \ii 2\bar{\rm
k'}z_j}.\la{78}\end{align}  
  With the help of the algebra  \eq{7}
we find (we recall that \((2\pi n_{\scriptscriptstyle\rm A})^{-1}=\nu\ell^2_\vv\))\begin{align}
\wick[offset=0.6em]{\c n_{ k}\c n_{ k'}}= \left(1-e^{\frac \nu 2{{ \ell_\vv^2\bm k}\cdot\bm { k'}}}\right)\no{  n_{ k+ k'}}.\la{81}
\end{align}
Then   \begin{align}\nn\begin{split}\wick[offset=0.9em]{\c\uu_z(  r)\c\uu_z( r')}&=-\Gamma^2(4\p_z\p_{z'})
\frac 1{V^2}\sum_{k,k'}e^{\ii\bm
k\bm r+\ii\bm k'\bm r'} \left(\frac{1-e^{\frac\nu 2\ell_\vv^2(\bm
k\cdot \bm k')}}{ k^2 k'^2}\right) \no{n_{ k+ k'}}.
\end{split}
\end{align}
Now we have to evaluate the expectation value \(\<n|\uu( r)\uu(r')| n\>\)  on the flow  (a quantum state) with the density \(n(r)\). But  we know
that  flows are the orbit of the Lie algebra \eq{15}, and therefore,  it is sufficient
to start from any state. The easiest is the ground state (the stationary
flow) \(|0\>\). At the uniform state we have \eq{202}. Hence,  
\begin{align}\begin{split}
\<0|\wick[offset=0.9em]{\c\uu_z({ r})\c \uu_z({ r}')}&|0\>=-\Gamma^2 (4\p_z\p_{z'})\cdot
\int e^{\ii{\bm
k}\cdot({\bm r}-{\bm r}')}\Big(\frac{1-e^{-\frac \nu 2{\ell^2_\vv k^2}}}{2\pi\ell_\vv^2k^4}\Big)\frac{\dd^2k}{(2\pi)^2}.\nn
\end{split}
\end{align}
 In the  limit, where points
are well separated, $k$ in the integral  is small and  the integrand behaves as $1/k^2$. At short distances  \(k\) is large, the integral converges. 
The crossover between two regimes is assisted by the factor $e^{-\frac \nu 2{\ell^2_\vv k^2}}$
 \begin{align}\begin{split}\la{311}
 &\<0|\wick[offset=0.9em]{\c \uu_z({ r})\c \uu_z({r}')}|0\>=-\frac\nu{2\pi}\Gamma^2(\p_z\p_{z'})\cdot
 \begin{cases}
               - \frac 1{2\pi}\log(|r-r'|/\sqrt V), & |r - r'|\gg \ell_\vv;\\
               - \frac 1{2\pi}\log(\ell_\vv/\sqrt{ V}) , & r\to r'.
            \end{cases}
            \end{split}
\end{align}
This is of course the Green function \(G_R\) of the Laplace operator regularized by the inter-vortex distance \(\ell_\vv\). We arrive to the expression for the quantum stress
 \begin{align}
T_{zz}=2(\hbar\Omega)\lim_{r\to r'}\partial_{z}
\partial_{z'}G_R({
r}, { r}').\la{251}
\end{align} 
We now  extend this result to a non-uniform flow. An economic way to do this is to invoke the geometric interpretation of the flow, where the vorticity is understood as a metric. Then, at a large distance (the upper line of \eq{311}) one expects to have the Green function of the Laplace-Beltrami operator \(\Delta=-(4/n)\p_z\p_{\bz}\) in the metric \(\dd s^2=n|\dd z|^2\).   At short distances the Green   function diverges as \(r\to r'\), but the result must be finite as in \eq{311}.  The covariant way to obtain the crossover is to remove the divergency at short distances  by  adding to the Green function  the logarithm of the geodesic distance \(d({
r}, { r}')\) between  points and then merge them
\begin{align}
G_R(r,r')\underset{r\to r'}{\approx}G({
r}, { r}')+\frac 1{2\pi}\log d({
r}, { r}').\la{R}
 \end{align}
 Then the infinity at short distances is subtracted in a covariant manner. This procedure yields a  regularized   Green function  consistent with the metric.    We may obtain the same result by   modifying   the factor $e^{-\frac \nu 2{\ell^2_\vv k^2}}$ in the integrand of \eq{R} by the operator $e^{\frac \nu 2{\ell^2_\vv \Delta}}$ and  replacing the integral by the trace.
 
 The result  of merging points in \eq{251}  is   known to be the Schwarzian of the metric \begin{align}\nn
\lim_{r\to r'}\partial_{z}
\partial_{z'}\Big[G({
r}, { r}')+\frac 1{2\pi}\log d({
r}, { r}')\Big]
=\frac{1}{24\pi} \left( \partial_z^{2}\log n-
\frac{1}{2}(\partial _z\log n)^{2}\right)&.\nn
\end{align}
Hence
 \begin{align}
T_{zz}=\frac{\hbar\Omega}{12\pi} \left( \partial_z^{2}\log n-
\frac{1}{2}(\partial_z\log n)^{2}\right).\la{35}
\end{align}
This effect is analogous to the gravitational anomaly in     quantum gravity. Physically, it is an   implementation a   flow dependent cut-off   \(\ell[n]\sim n^{-1/2}\). 

\section{Trace Anomaly }
There is a subtle distinction between the stress acting on the irrotational  flow around vortices and forces exerted on vortices \cite{bogatskiy2019vortex}. There is no  difference in traceless part of the stress, shear forces acting on the fluid and vortex cores are the same. However, the trace of the stress supported by the flow and the vortex cores are different. The former is the pressure determined by the incompressibility condition. The latter is  the   trace of  the quantum stress \(T^{\bz}_z\).   Since vorticity is the metric the stress tensor with respect to the metric    is divergence-free. In complex coordinates, where the metric is \(\dd s^2=n|\dd z|^2\) the divergence-free condition reads  \begin{align}
\p_{\bz} T_{zz}+n\p_z (n^{-1}T_{z\bz})=0,\quad T^{\bz}_z=n^{-1}T_{z\bz}.   \la{31}
\end{align} 
With the help of \eq{35} and \eq{31} we obtain the trace  \begin{align}
 T^{\bz}_z=\frac{\hbar\Omega}{48\pi}
\mathcal{R}.\la{36}
\end{align}
This formula  often refers as a trace anomaly: the trace
of the quantum stress
is proportional to the  curvature  \eq{101}. 

However,   in  the metric of the host surface \(|\dd z|^2\) the stress  tensor does not conserve. As it follows from
\eq{31} 
\begin{align}\la{37}
\nabla^iT_{ij}=\12T^{\bz}_z\,\nabla_j  n.
\end{align}
(we use covariant derivatives with respect to  the  host surface). 
This is the source of the quantum correction to the Helmholtz equation.
\section{ Quantum corrections to the Helmholtz law} The quantum Helmholtz equation follows from  (\ref{37},\ref{36}) and \eq{27}. It reads
 \begin{align}\begin{split}
\la{40}
D_t\omega&=\frac {\hbar}
{ m_\vv}\, \, \frac 1{96\pi}\nabla \mathcal{R\times}\nabla\omega. 
\end{split}
\end{align}
In terms of the  Euler equation correction  reads
 \begin{align}\begin{split}
\nn
D_t u+\nabla p&=\frac {\hbar}
{ m_\vv}\, \, \frac 1{48\pi} \mathcal{R\times}\Delta u. 
\end{split}
\end{align}
In this equation  all terms are normally ordered, hence it  must be treated as a classical equation. Since vorticity
is a metric, this equation can be interpreted geometrically
as a metric flow. 

We also comment that this equation \eq{40} is Hamiltonian. The quantum correction amounts
the addition to  the   Hamiltonian the term
\begin{align}
H\to H+\nu\frac{\hbar\Omega}{96\pi} \int (\nabla \log n)^2 \dd V.
\end{align} 
Clearly, this correction is the leading term of the gradient expansion. The omitted terms are sensitive to the regularization of vortex cores. They are not controlled by any symmetry.

Other variables frozen into the classical flow obey the same equation. For example, canonical   Clebsch variables  defined by \eq{Cl} 
flow according to the equation similar to   \eq{40}\begin{align}
 \rho_{\scriptscriptstyle\rm A}D_t\lambda^a=\nabla T_z^{\bz}\times\nabla\lambda^a.
\end{align} If we treat the curvature as a function of Clebsch's coordinates,  the Helmholtz equation appears in the canonical form\begin{align}
 \rho_{\scriptscriptstyle\rm A}D_t\lambda^a=-\omega\epsilon^{ab}\frac{\partial
 {T_z^{\bz}}}{\partial\lambda^b}.\quad 
\end{align} 

If vorticity is a small  wave   \(\omega= 2\Omega+V^{-1}  \sum_{k\neq 0} e^{\ii\bm k\cdot\bm r} \omega_k\)  we  may expand the quantum Helmholtz equation  \eq{40} about the net vorticity. In the  harmonic approximation (the leading order in  \(\omega_k\)) we obtain 
%
 \begin{align}
 n_\vv  D_t {\omega_k}=   \frac 1V\sum_q \frac{q^2} {96\pi}e_{qk}\,{\omega_{q}\omega_{k-q}},\la{222}
  \end{align}
where $e_{qk} $ are the structure constants of $\bf SDiff$ \eq{172} (compare
with \cite{zakharov1989algebra})
. In  some physical applications the sum over modes  could be truncated, but vorticity waves on the uniform background are never linear.

Eqs.\eq{40} and \eq{222}
are our main result.  

\section{Virasoro algebra} 
 We  briefly comment on the relation of the  quantum stress and  the  Virasoro algebra. To shorten the formulas we set \(\hbar\Omega=1\).


Since, a change of vorticity \(n\to n+\delta n\) is the Weyl transformation of the  metric, the trace of the quantum  stress  is a generator of dilatations. For an arbitrary operator   \(\mathcal{O}\) we have  \begin{align}\begin{split}\nn & \no{n\frac\delta{\delta n}\mathcal{O}}=
-\wick[offset=1.0em]{\c T_{z\bz}\c{  \mathcal{O}}}.\end{split}\end {align}
 With the help of  the conservation law \eq{31} and the \(\bar\p\) - formula we  express this relation through the traceless part of the stress  \begin{align}
\frac{1}\pi \int\frac{n(\xi)}{z-\xi}\p_\xi\left[\frac{\delta\mathcal{O}}{\delta n(\xi)}\right]\dd^2 \xi=\wick[offset=1.0em]{\c T_{zz}\c{  \mathcal{O}}}\nn
\end{align} 
(we  dropped the normal ordering symbol).

We now specify  \(\mathcal{O}={T}_{zz}(z')\tilde{\mathcal{O}}\) and assume that \(\tilde{\mathcal{O}}\) does not depend on \(n\) \begin{align}
\frac{1}\pi \int\frac{n(\xi)}{z-\xi}\p_\xi\no{\left[\frac{\delta{T}_{zz}(z')}{\delta
n(\xi)}\right]\dd^2 \xi\, \tilde{\mathcal{O}}}=\no{T_{zz}(z)T_{zz}(z')\tilde {\mathcal{O}}}-T_{zz}(z)\no{T_{zz}(z')\tilde{\mathcal{O}}}. \la{47}
\end{align}We now evaluate the lhs  of  \eq{47} with the help of \eq{35} on  a stationary flow by setting \(n=n_\vv\) in the result.  Eq. \eq{47} is equivalent to  
the conformal  Ward identity, hence the calculations are standard, see, e.g., \cite{belavin1984,eguchi1987}. The result is well known\begin{align} \nn
 & \no{T_{zz}(z)T_{zz}(z')\tilde {\mathcal{O}}}-T_{zz}(z)\no{T_{zz}(z')\tilde{\mathcal{O}}}
 =\frac{1/2}{(z-z')^2}\no{ \tilde{\mathcal{O}}}-\left(\frac 2{(z-z')^2}+\frac 1{z-z'}\p_{z'}\right) \no{T_{zz}(z')\tilde{\mathcal{O}}}\nn
\end{align}
This relation is equivalent to the Virasoro algebra, whose generators \(L_n\) are defined by the Laurent  series \(T_{zz}(z)=- \sum_n (z-z')^{-n-2}L_n\)
about a point \(z'\) (notice the minus sign in the definition)
\begin{align}
[L_n,\ L_m]=(n-m)L_{n+m}+\frac c{12}(n^3-n)\delta_{n+m,0}.\nn
\end{align}
 The   central extension of the Virasoro algebra appears to be \(c=1\).
 
\section{Quadrupole modes}
We conclude by a brief description of the global symmetry of the quantum flow.

The infinite dimensional algebra \textbf{SDiff} possesses  \(sl(2,\mathbb{R)} \)    subalgebra (isomorphic
to \(su(1,1)\approx sp(2,\mathbb{R})\approx so(2,1)\)). 
The maximal compact subgroup of  \(SL(2,\mathbb{R})\)   is \(SO(2)\),   a group of planar rotations, the global symmetry of the fluid. The generator  of   \(SO(2)\) is the angular impulse \eq{171} \(L=m_\vv\sum_i(\bm r_i\times\bm
v_i\)).  Therefore, the Hamiltonian and the  angular impulse act  as diagonal operators in a module   \(\left(SL(2,\mathbb{R}),\ SO(2)\right)\).   We use its eigenvalues of the angular impulse 
to label weight states \(L|l\>=\hbar l|l\>\).   

  In geometry of  a (punctured) sphere we can realize  \(sl(2,\mathbb{R)} \) algebra   by quadrupole moment of vorticity, the rank-2 symmetric  tensor \begin{align}\nn
Q_{ij}=-\frac{\rho_{\scriptscriptstyle\rm A}}2\int  (x_i x_j)\,\omega \dd V,\quad 
\end{align} where the integration extends on the entire fluid.  In complex coordinates \begin{align}\begin{split}\nn
&Q_{\bz\bz}=-\frac{\rho_{\scriptscriptstyle\rm A}}2\int z^2\omega \dd V={\pi \hbar n_{\scriptscriptstyle\rm A}} \sum_i
z_i^2,\\ 
& Q_{z\bz}=-\frac{\rho_{\scriptscriptstyle\rm A}}2\int  |z|^2\,\omega
\dd V=\pi\hbar{ n_{\scriptscriptstyle\rm A}} \sum_i
\12(z_iz_i^\dag+z_i^\dag z_i).
\end{split}
\end{align}   The Lie-Poisson algebra of quadrupoles follows from \eq{6}
 \begin{align}\nn
& \{Q_{ij},\  Q_{kl}\}=\epsilon_{ik}Q_{jl}+\epsilon_{jk}Q_{il}+\epsilon_{il}Q_{jk}+\epsilon_{jl}Q_{ik}.
\end{align}
When we replace the 
Poisson brackets  by the commutator
we obtain   \(sl(2,\mathbb{R)} \) Lie algebra. In complex coordinates it reads
\begin{align}\nn\begin{split}
    [Q_{z\bz},\  Q_{zz}]=-&\hbar Q_{zz},\quad [Q_{z\bz},\  Q_{\bz\bz}]=\hbar
Q_{\bz\bz},\\&[Q_{\bz\bz},\  Q_{zz}]=2\hbar Q_{z\bz}.\end{split}\end{align}
       The  trace of the quadrupole moment     is  conserved \(\frac{\dd}{\dd t} Q_{z\bz}=(\ii/\hbar)[H,Q_{z\bz}]=0\)
 \cite{saffman1992vortex}.  This is  an immediate consequence of the Kirchhoff or Euler equations. Up to an additive number which  depends on the choice of   a  frame,  and an overall factor,
\(Q_{z\bz}\) is  the angular impulse \(L\).    The  eigenvalues of  
\(Q_{z\bz}\)  follow from    the Kirchhoff equation  \eq{4}\begin{align}\nn
Q_{z\bz}|l\>=\hbar\left( k+\frac l 2\right)|l\>,
\end{align}
where  \(k=\frac18N_{\scriptscriptstyle\rm A}(N_\vv-1)\). This constant is called Bargmann index.
It gives the value to the  Casimir operator of   \(sl(2,\mathbb{R)} \)
\be C_2=Q_{z\bz}^2-\12(Q_{zz}Q_{\bz\bz}+Q_{\bz\bz}Q_{zz})=\hbar^2k(k-1).\ee
 
 The  operators   \(Q_{\bz\bz}\) and \(Q_{zz}\) generate clockwise and anticlockwise shear flows. They are ladder operators changing  the angular
impulse by \(\pm2\hbar\) 
\begin{align}
Q_{zz}|l\>=-\frac {\hbar l}4 |l-2\>, \quad Q_{\bz\bz}|l\>=\hbar( k+\frac {  l}4 )|l+2\>
.\nn
\end{align}    Therefore
the  flow with a given angular impulse is the weight state   raised and   lowered by   shear-quadrupole operators by the increment 2. Say, a shear operator \(Q_{\bz\bz}\) acting on the ground state \(|0\>\) creates a  spin-2 state   (called squeezed modes in optics). Furthermore, the   ``squeezing operator'' \(\exp{({\bar\xi}^2Q_{\bz\bz}-
\xi^2 Q_{zz})}\) creates a flow corresponding to     \(sl(2,\mathbb{R)} \) coherent state \cite{perelomov}.

 Let us now turn to the dynamics of   the quadrupole  operators. From Kirchhoff equation we obtain   \be\nn \ii\frac{\dd}{\dd t} Q_{zz}= \frac{\hbar\Omega}\nu\sum_{i> j}2 e^{-2\ii\theta_{ij}},\ee where \(\theta_{ij}={\rm arg}(z_i-z_i)\). The coarse-grained version  of this expression    follows from
 \eq{27}  
  \begin{align}\la{28}\begin{split}
  D_tQ_{zz}=\frac \ii 2\int
  T_{zz}\dd V,
\end{split}
\end{align}
where \(D_tQ_{zz}=\dot Q_{zz}-\frac \ii 2\int
  \uu_{z}^2\dd V\) includes the  advection.  The rhs is the effect of the quantum stress. In classical
fluid only   advection deforms the  quadrupole moment of vorticity. For example,  the  viscous stress does not contribute to the volume integral of the rhs. However, the quantum stress does (see \eq{271}).
  Its effect stabilizes instabilities driven by the advection.
In a near stationary flow with a small advection, the quantum stress could be a dominant contribution.
If the bulk vorticity is uniform the correction reads \be D_tQ_{zz}=\frac  {1}{2\ii } \hbar\Omega V L_{-2}.\ee 
The rate of change of the uniform shear is proportional to  dilatation.
 
\bigskip
\bigskip


Summing up, we computed the quantum correction to the  Euler equation for 2D chiral flows.   The quantum corrections appears in the form of the quantum stress, an additional force exerted on quantum vortices. The quantum stress consists of higher derivatives and  possesses a scale  of fundamental vortex circulation \(\Gamma\).  It destroys the scaling invariance of the Euler equation. The implication of the quantum stress is that the diagonal matrix elements of vorticity are no longer frozen into the
flow.  This is the price of maintaining a   fundamental symmetry of fluid flows,  the relabeling symmetry. The Kelvin circulation theorem, however, remains intact.
Vice versa, the quantum corrections are unambiguously fixed by the relabeling symmetry. This effect is  parallel to the quantization of 2D gravity, where the invariance with respect to diffeomorphisms uniquely determines the Polyakov-Liouville action. In both cases upholding the diffeomorphism invariance is implemented through the gravitational anomaly.
There is no surprise that the Lie-Poisson algebra of the stress is the Virasoro algebra, the centrally extended algebra of holomorphic diffeomorphisms. 
 \bigskip
 
  The author acknowledges helpful discussion with M.F. Lapa.

\bibliographystyle{unsrtnat}
\bibliography{FQHE_ref}

\begin{thebibliography}{27}
\providecommand{\natexlab}[1]{#1}
\providecommand{\url}[1]{\texttt{#1}}
\expandafter\ifx\csname urlstyle\endcsname\relax
  \providecommand{\doi}[1]{doi: #1}\else
  \providecommand{\doi}{doi: \begingroup \urlstyle{rm}\Url}\fi

\bibitem[Khalatnikov(1989)]{Khalatnikov}
I.~M. Khalatnikov.
\newblock \emph{An introduction to the theory of superfluidity}, volume~23.
\newblock Perseus Books, 1989.

\bibitem[Wiegmann(2013{\natexlab{a}})]{wiegmann2013hydrodynamics}
P.~B. Wiegmann.
\newblock Hydrodynamics of {E}uler incompressible fluid and the fractional
  quantum {H}all effect.
\newblock \emph{Phys. Rev. B}, 88:\penalty0 241305, 2013{\natexlab{a}}.
\newblock \doi{10.1103/PhysRevB.88.241305}.
\newblock URL \url{https://link.aps.org/doi/10.1103/PhysRevB.88.241305}.

\bibitem[Wiegmann(2013{\natexlab{b}})]{wiegmann2013JETP}
P.~Wiegmann.
\newblock Anomalous hydrodynamics of fractional quantum {H}all states.
\newblock \emph{J.Exp.Theor.Phys.}, 117\penalty0 (3):\penalty0 538--550,
  2013{\natexlab{b}}.
\newblock \doi{10.1134/s1063776113110162}.
\newblock URL \url{https://doi.org/10.1134%2Fs1063776113110162}.

\bibitem[Arnold and Khesin(1999)]{arnold1999topological}
V.~I. Arnold and B.~A Khesin.
\newblock \emph{Topological methods in hydrodynamics}, volume 125.
\newblock Springer, 1999.

\bibitem[Landau(1941)]{Landau1941}
L.~D. Landau.
\newblock Theory of superfluidity of helium {II}.
\newblock \emph{J. Phys.(USSR)}, 5:\penalty0 71, 1941.

\bibitem[Saffman(1992)]{saffman1992vortex}
P.~G Saffman.
\newblock \emph{Vortex dynamics}.
\newblock Cambridge university press, 1992.

\bibitem[Girvin et~al.(1986)Girvin, MacDonald, and Platzman]{girvin1986}
S.~M. Girvin, A.~H. MacDonald, and P.~M. Platzman.
\newblock Magneto-roton theory of collective excitations in the fractional
  quantum {H}all effect.
\newblock \emph{Phys. Rev. B}, 33\penalty0 (4):\penalty0 2481--2494, 1986.
\newblock \doi{10.1103/physrevb.33.2481}.
\newblock URL \url{https://doi.org/10.1103%2Fphysrevb.33.2481}.

\bibitem[Bargmann(1961)]{bargmann1961}
V.~Bargmann.
\newblock On a {H}ilbert space of analytic functions and an associated integral
  transform.
\newblock \emph{Communications on Pure and Applied Mathematics}, 14\penalty0
  (3):\penalty0 187--214, 1961.
\newblock \doi{10.1002/cpa.3160140303}.
\newblock URL \url{https://doi.org/10.1002%2Fcpa.3160140303}.

\bibitem[Laughlin(1983)]{Laughlin1983}
R.~B. Laughlin.
\newblock {Anomalous Quantum {H}all Effect: An Incompressible Quantum Fluid
  with Fractionally Charged Excitations}.
\newblock \emph{Phys. Rev. Lett.}, 50:\penalty0 1395--1398, 1983.
\newblock \doi{10.1103/PhysRevLett.50.1395}.
\newblock URL \url{http://link.aps.org/doi/10.1103/PhysRevLett.50.1395}.

\bibitem[Zabrodin and Wiegmann(2006)]{Zabrodin2006}
A~Zabrodin and P~Wiegmann.
\newblock Large-nexpansion for the {2D} {D}yson gas.
\newblock \emph{Journal of Physics A: Mathematical and General}, 39\penalty0
  (28):\penalty0 8933--8963, 2006.
\newblock \doi{10.1088/0305-4470/39/28/s10}.
\newblock URL \url{https://doi.org/10.1088%2F0305-4470%2F39%2F28%2Fs10}.

\bibitem[Feynman(1955)]{FEYNMAN1955}
R.P. Feynman.
\newblock Application of quantum mechanics to liquid helium.
\newblock \emph{Progress in Low Temperature Physics}, 1:\penalty0 17 -- 53,
  1955.
\newblock ISSN 0079-6417.
\newblock \doi{http://dx.doi.org/10.1016/S0079-6417(08)60077-3}.
\newblock URL
  \url{http://www.sciencedirect.com/science/article/pii/S0079641708600773}.

\bibitem[Fairlie and Zachos(1989)]{fairlie1989infinite}
D.~B. Fairlie and C.~K. Zachos.
\newblock Infinite-dimensional algebras, sine brackets, and {S}{U}({N}).
\newblock \emph{Physics Letters B}, 224\penalty0 (1-2):\penalty0 101--107,
  1989.

\bibitem[Polyakov(1987)]{Polyakov1987}
A.~M. Polyakov.
\newblock \emph{Gauge Fields and Strings (Contemporary Concepts in Physics)}.
\newblock Harwood Academic Publishers, Switzerland, 1987.

\bibitem[Bekarevich and Khalatnikov(1961)]{bekarevich}
I.~L. Bekarevich and I.~M. Khalatnikov.
\newblock Phenomenological derivation of the equations of vortex motion in {H}e
  {II}.
\newblock \emph{Sov. Phys. JETP}, 13:\penalty0 643, 1961.

\bibitem[Kemoklidze and Khalatnikov(1964)]{kemoklidze}
M.~P. Kemoklidze and I.~M. Khalatnikov.
\newblock Hydrodynamics of rotating helium {II} in an annular channel.
\newblock \emph{JETP}, 19:\penalty0 1134, 1964.
\newblock URL \url{http://www.jetp.ac.ru/cgi-bin/dn/e_019_05_1134.pdf}.

\bibitem[Hall(1960)]{hall1960}
H.~E. Hall.
\newblock The rotation of liquid helium {II}.
\newblock \emph{Advances in Physics}, 9\penalty0 (33):\penalty0 89--146, 1960.

\bibitem[Hall and Vinen(1956)]{hall-vinen}
H.~E. Hall and W.~F. Vinen.
\newblock The rotation of liquid helium {II}, {II}. {T}he theory of mutual
  friction in uniformly rotating helium {II}.
\newblock \emph{Proc R Soc Lond A Math Phys Sci}, 238\penalty0 (1213):\penalty0
  215--234, 1956.

\bibitem[Wiegmann and Abanov(2014)]{AW}
P.~Wiegmann and A.~G. Abanov.
\newblock Anomalous hydrodynamics of two-dimensional vortex fluids.
\newblock \emph{Phys. Rev. Lett.}, 113:\penalty0 034501, 2014.
\newblock \doi{10.1103/PhysRevLett.113.034501}.
\newblock URL \url{http://link.aps.org/doi/10.1103/PhysRevLett.113.034501}.

\bibitem[Bogatskiy(2019)]{bogatskiy2019vortex}
A.~Bogatskiy.
\newblock Vortex flows on closed surfaces.
\newblock \emph{arXiv:1903.07607}, 2019.
\newblock URL \url{https://arxiv.org/pdf/1903.07607.pdf}.

\bibitem[Bogatskiy and Wiegmann(2019)]{bogatskiyedge}
A.~Bogatskiy and P.~Wiegmann.
\newblock Edge wave and boundary layer of vortex matter.
\newblock \emph{Phys. Rev. Lett.}, 122:\penalty0 214505, 2019.
\newblock \doi{10.1103/PhysRevLett.122.214505}.
\newblock URL \url{https://link.aps.org/doi/10.1103/PhysRevLett.122.214505}.

\bibitem[Landau and Lifshitz(1987)]{Landau6}
LD~Landau and EM~Lifshitz.
\newblock \emph{Fluid Mechanics: Volume 6 (Course of Theoretical Physics S).}
\newblock Pergamon, 1987.

\bibitem[Andreev and Kagan(1984)]{andreev1984}
A.~F. Andreev and M.~Yu. Kagan.
\newblock Hydrodynamics of a rotating superfluid liquid.
\newblock \emph{Zh. Eksp. Teor. Fiz}, 86:\penalty0 546--557, 1984.

\bibitem[Volovik and Dotsenko(1980)]{volovik1980}
G.~E. Volovik and V.~S. Dotsenko.
\newblock Hydrodynamics of defects in condensed media, using as examples
  vortices in rotating {H}e {I}{I} and disclinations in a planar magnet.
\newblock \emph{Soviet Phys. JETP}, 78\penalty0 (1):\penalty0 132--148, 1980.

\bibitem[Zakharov(1989)]{zakharov1989algebra}
Vladimir~Evgen'evich Zakharov.
\newblock The algebra of integrals of motion of two-dimensional hydrodynamics
  in clebsch variables.
\newblock \emph{Functional Analysis and Its Applications}, 23\penalty0
  (3):\penalty0 189--196, 1989.

\bibitem[Belavin et~al.(1984)Belavin, Polyakov, and Zamolodchikov]{belavin1984}
A.~A Belavin, A.~M Polyakov, and A.~B Zamolodchikov.
\newblock Infinite conformal symmetry in two-dimensional quantum field theory.
\newblock \emph{Nuclear Physics B}, 241\penalty0 (2):\penalty0 333--380, 1984.

\bibitem[Eguchi and Ooguri(1987)]{eguchi1987}
T.~Eguchi and H.~Ooguri.
\newblock Conformal and current algebras on a general {R}iemann surface.
\newblock \emph{Nuclear Physics B}, 282:\penalty0 308--328, 1987.

\bibitem[Perelomov(2012)]{perelomov}
A.~Perelomov.
\newblock \emph{Generalized coherent states and their applications}.
\newblock Springer, 2012.

\end{thebibliography}
\end{document}